\documentclass[11pt,a4paper]{article}
\usepackage{amsmath,amssymb}
\usepackage{epsfig,graphicx}
\usepackage{subfigure}
\usepackage{graphicx}
\usepackage{rotating}
\usepackage{bm}
\usepackage{axodraw}
\usepackage{color}
\usepackage[sort&compress,square]{natbib}

\def\M{{{\cal M}}}

\def\Tr{{\rm Tr}}

\def\det{{\rm det}}

\def\Dbarslash{\,\,{\raise.15ex\hbox{/}\mkern-12mu {\bar\D}}}
\def\Dslash{\,\,{\raise.15ex\hbox{/}\mkern-12mu \D}}
\def\delslash{\,\,{\raise.15ex\hbox{/}\mkern-9mu \partial}}
\def\delbarslash{\,\,{\raise.15ex\hbox{/}\mkern-9mu {\bar\partial}}}

\let\vev=\Vev

\def\D{{\cal D}}
\def\Dbarslash{\,\,{\raise.15ex\hbox{/}\mkern-12mu {\bar\D}}}
\def\delslash{\,\,{\raise.15ex\hbox{/}\mkern-9mu \partial}}
\def\Dslash{\,\,{\raise.15ex\hbox{/}\mkern-12mu \D}}

\def\={\, =\, }
\def\+{\, +\, }
\def\-{\, -\, }

\newcommand{\be}{\begin{equation}}
\newcommand{\ee}{\end{equation}}
\def\bea{\begin{eqnarray}}
\def\eea{\end{eqnarray}}

\begin{document}
\date{\mbox{ }}
\title{{\normalsize  IPPP/07/39; DCPT/07/78\hfill\mbox{}\hfill\mbox{}}\\
\vspace{2.5cm} \Large{\textbf{Dynamical breaking of $U(1)_{R}$ and supersymmetry \linebreak in a metastable
vacuum}}}
\author{Steven Abel, Callum Durnford, Joerg Jaeckel, Valentin V. Khoze\\[2ex]
\small{\em Institute for Particle Physics Phenomenology,}\\
\small{\em Durham University, Durham DH1 3LE, United Kingdom} }
\date{}
\maketitle

\begin{abstract}
\noindent
We consider the metastable $\mathcal{N}=1$ QCD model of Intriligator,
Seiberg and Shih (ISS), deformed by adding a baryon term to the superpotential.
This simple deformation causes the spontaneous breaking
of the approximate $R$-symmetry of the metastable vacuum.
We then gauge the flavour $SU(5)_{f}$ and identify it with the parent gauge symmetry of the Standard Model (SM).
This implements direct mediation of supersymmetry breaking without the need for an additional messenger
sector. A reasonable choice of parameters
leads to gaugino masses of the right order. Finally, we speculate that the entire ``ISS $\times$ SM'' model should be interpreted as
a magnetic dual of an (unknown) asymptotically free theory.
\end{abstract}

\vspace{3ex}

The issue of supersymmetry (SUSY) breaking has recently been reinvigorated
by Intriligator Seiberg and Shih \cite{ISS} (ISS). Their observation
-- that metastable SUSY breaking vacua can arise naturally and dynamically
in the low-energy limit of supersymmetric $SU(N)$ gauge theories
-- has important implications for our understanding of how SUSY is
broken in nature. Following this work there has been exploration of
both the cosmological consequences \cite{ACJK,heat2,heat3,heat4,heat5},
and the possibilities for gauge or direct mediation of the SUSY breaking
to the visible sector \cite{ISS,Forste:2006zc,Amariti:2006vk,Dine:2006gm,DM,Kitano:2006xg,MN,AS,Csaki:2006wi,Abel:2007uq,
Amariti:2007qu,Murayama:2007fe,Intriligator:2007py,Intriligator:2007cp,Abel:2007zm,Shih:2007av,Ferretti:2007ec,Cho:2007yn,
Ooguri:2007iu,Katz:2007gv,Brummer:2007ns,Dudas:2007hq,Delgado:2007rz}.
On the cosmological side, the attractive feature of these models is
that the metastable vacua are naturally long lived due to the flatness
of the potential. Moreover, at high temperatures the SUSY breaking
vacua are dynamically favoured over the SUSY preserving ones because
they have more light degrees of freedom, so the early Universe would
naturally have been driven into them%
\footnote{An additional cosmological development which is somewhat orthogonal
to our discussion is the use of the models of ISS in supergravity
to ``uplift'' supersymmetric models to small non-zero cosmological
constant \cite{Dudas:2006gr,Kallosh:2006dv,Lebedev:2006qc}. This
possible phenomenon is relevant for larger SUSY breaking than that
we will be considering here.%
}. On the phenomenological side, attention has focussed on a striking
aspect of metastability, namely that the models do not have an exact
$U(1)_{R}$-symmetry, and indeed the $U(1)_{R}$-symmetry is anomalous
under the same gauge group that dynamically restores the supersymmetry
in the supersymmetric global minima.

\emph{In principle} this allows one to evade the theorem by Nelson
and Seiberg that SUSY breaking requires $R$-symmetry in a generic
model (i.e. one that includes all couplings compatible with the symmetries)
\cite{Nelson:1993nf}. $R$-symmetry is unwelcome because it implies
that gauginos are massless, so the fact that it can be broken by metastability
is an encouraging sign. In a more recent paper \cite{Intriligator:2007py}
it was emphasized that the relation between SUSY breaking and $R$-symmetry
is a continuous one, in the sense that the lifetime of the metastable
vacuum decreases in proportion to the size of any explicit $R$-symmetry
breaking terms that one adds to the theory. This allows one to play
the {}``approximate $R$-symmetry'' game: add to the superpotential
of the effective theory explicit $R$-symmetry breaking terms of your
choosing, whilst trying to keep the metastable minimum as stable as
possible. Such approximate $R$-symmetry models can then be motivated
by appealing to (for example) higher dimensional operators of the
underlying high energy physics.

Clearly there is some tension in this procedure. For example the gauge
mediation scenario explored in refs.\cite{MN,Murayama:2007fe} invokes
a messenger sector (denoted by $f$). The field $f$ has to have an explicit
$R$-breaking mass-term to give gauginos a mass, and consequently
a new SUSY restoring direction opens up along which $f$ gets a vev.
One is then performing a rather delicate balancing act: in order to
avoid disastrously fast decay of the metastable vacuum, large SUSY
breaking scales must be invoked so that the $R$-breaking mass can
be sufficiently small. It should also be noted that here the $R$-symmetry
breaking responsible for the globally supersymmetric minima of ISS
models plays no direct role in the generation of gaugino masses, and
consequently this is expected to be a generic problem for gauge mediation
of metastable SUSY breaking. This is also a problem for the models
that were constructed to implement direct mediation \cite{Csaki:2006wi},
and again, in those cases certain operators had to be forbidden by
hand, making the superpotential non-generic.

To avoid these problems, the next option for generating non-zero gaugino
masses would be to use the explicit $R$-breaking of the ISS model
itself, associated with the metastability and the existence of a global
supersymmetric groundstate. This is in fact a more difficult proposition
than one might suppose for the following reason. At the metastable
minimum there is an unbroken approximate $R$-symmetry (which is of
course why it is metastable in the first place). The $R$-symmetry is explicitly (more precisely anomalously) broken only by the nonperturbative term,
\begin{equation}
\label{nonperturbative}
W_{np}\propto(\det\,_{N_{f}}\Phi)^{\frac{1}{N}}\sim\Phi^{\frac{N_{f}}{N}},
\end{equation}
where $\Phi$ is the meson field, $SU(N)_{mg}$ is the gauge group of the magnetic theory, and $N=N_{f}-N_{c}$
with $SU(N_{c})$ being the gauge group of the electric theory \cite{ISS}. One
might hope that this would induce (for example) $R$-symmetry breaking
mass-terms that contribute to gaugino masses in perturbation theory.
However such mass-terms will be typically of order $\frac{\partial^{2}W}{\partial\Phi^{2}}\sim\Phi^{\frac{2N_{c}-N_{f}}{N}}$.
Thus since ISS models are valid in the interval $N_{c}+1\leq N_{f}<\frac{3}{2}N_{c}$,
they are exactly zero in the metastable minimum where $\langle\Phi\rangle=0$.

We are led to an alternative -- the focus of this paper -- which is
to \emph{spontaneously} break the approximate $R$-symmetry of the
ISS model to generate gaugino masses. The explicit breaking of the
model then ensures that any $R$-axions get a mass and are made safe. The
natural avenue to explore is to gauge (part of) the $SU(N_{f})$ flavour
symmetry of the ISS model, identifying it with the Standard Model
gauge groups. This would allow the quarks and mesons in the theory
to mediate the SUSY breaking directly to the Standard Model, thereby
avoiding the need for any messenger sectors which as we have seen
are liable to destabilize the metastable vacuum. Once spontaneous
$R$-breaking has been achieved, there is in principle nothing to
prevent it being mediated via these fields to the rest of the model
including the gaugino masses %
\footnote{Indeed the very same point was made in Ref.~\cite{DM} which was presented
in the language of retrofitting. There however, successful mediation
required a messenger sector which, in general, may lead to new and unstable directions. %
}. (Note that the nonperturbative explicit $R$-breaking can also now
contribute to gaugino masses since $\Phi$ will get a vev.)

In this paper we demonstrate that perfectly viable direct mediation
of SUSY breaking can indeed be implemented in this way, by making
the simplest deformation to the ISS model that one can imagine, namely
the addition of a baryon term to the superpotential. This {}``baryon-deformed'' QCD model has
a runaway direction to a non-supersymmetric metastable minimum of the ISS type, along
a particular direction of field space which is lifted by the Coleman-Weinberg
potential and stabilized. Along this direction the meson modes $\Phi_{ij}$
acquire a vev, and the approximate $R$-symmetry is spontaneously
broken. Importantly the diagonal ($U(1)$-trace) component of the
pseudo-Goldstone modes (i.e. those modes of $\Phi_{ij}$ whose flavour
indices correspond to SM gauge group) acquires a vev at this point
as well; the latter gives $R$-breaking masses to the magnetic quarks
which are charged under the SM gauge groups. This enables them to
act as messenger fields giving the gauginos masses at one-loop. We
stress that all of this happens automatically upon adding a baryon.
There is no need for any messenger sector outside the ISS model, and
therefore no additional instability is induced. Moreover, we will
show that the resulting gaugino masses can be naturally
of the right order.\\

Let us begin by introducing our model which is based on the model of
ISS with $SU(N_{c})$ gauge symmetry and $N_{f}$ flavours of
quark/anti-quark pairs in the electric theory. The magnetic dual
theory has a $SU(N)_{mg}$ gauge symmetry, where $N=N_{f}-N_{c}$,
$N_{f}$ flavours of fundamental quark/anti-quark pairs, and is IR
free if $N_{c}+1\leq N_{f}<\frac{3}{2}N_{c}$. The minimal values
consistent with this equation and leading to a non-trivial magnetic
gauge group are $N_{f}=7$ and $N_{c}=5$ giving
$SU(2)_{mg}$ in the magnetic dual theory. Now consider the following
superpotential:
\begin{equation}
W=\Phi_{ij}\varphi_{i}.\tilde{\varphi_{j}}-\mu_{ij}^{2}\Phi_{ji}+m\varepsilon_{ab}\varepsilon_{rs}\varphi_{r}^{a}\varphi_{s}^{b}\,\,\end{equation}
where $i,j=1...7$ are flavour indices, $r,s=1,2$ run over the first
two flavours only, and $a,b$ are $SU(2)_{mg}$ indices (we set the coupling
$h=1$ for simplicity). This is the superpotential of ISS with the
exception of the last term which is a baryon of $SU(2)_{mg}$. Note that
the 1,2 flavour indices and the 3...7 indices have a different status
and the flavour symmetry is broken explicitly to $SU(2)_{f}\times SU(5)_{f}$.
The $SU(5)_{f}$ factor will be gauged separately and will be identified
with the parent $SU(5)$ of the SM%
\footnote{Note that the breaking of $SU(5)$ is assumed to take place or be
included explicitly in the SM sector.%
}.

The baryon deformation is the leading order deformation of
the ISS model that is allowed by R-symmetry (as well as the
gauge and flavor symmetries discussed above). Terms quadratic in the
mesons that could arise from lower dimensional irrelevant operators
in the electric theory are forbidden by R-symmetry.

Using the $SU(2)_{f}\times SU(5)_{f}$ symmetry, the matrix
$\mu_{ij}^{2}$ can be brought to a diagonal form \begin{equation}
\mu_{ij}^{2}=\left(\begin{array}{cc}
\mu^{2}\mathbf{I}_{2} & 0\\
0 & \hat{\mu}^{2}\mathbf{I}_{5}\end{array}\right).\end{equation}
We will assume that $\mu^{2}>\hat{\mu}^{2}$. The parameters $\mu^{2}$,
$\hat{\mu}^{2}$ and $m$ have an interpretation in terms of the electric
theory: $\mu^{2}\sim\Lambda m_{Q}$ and $\hat{\mu}^{2}\sim\Lambda\hat{m}_{Q}$
come from the electric quark masses $m_{Q}$, $\hat{m}_{Q}$, where
$\Lambda$ is the Landau pole of the theory. The baryon operator can
be identified with a corresponding operator in the electric theory.
Indeed the mapping from baryons $B_{E}$ in the electric theory to
baryons $B_{M}$ of the magnetic theory, is $B_{M}\Lambda^{-N}\leftrightarrow B_{E}\Lambda^{-N_{c}}$
(we neglect factors of order one). Thus one expects
\begin{equation}
m\sim M\left(\frac{\Lambda}{M}\right)^{2N_{c}-N_{f}}=\frac{\Lambda^{3}}{M^{2}}.
\end{equation}
Where $M$ represents the scale of new physics in the electric
theory at which the irrelevant operator $B_{M}$ is generated. We
will think of it as being $M_{P}$ or $M_{GUT}$ although as we shall
see a large range of values can be accommodated.

It is encouraging that this rather minimal choice of parameters allows
us to identify $SU(5)_{f}$ flavour symmetry with the Standard Model
gauge groups\footnote{It is also an amusing coincidence that the electric theory has the same gauge
groups for colour and flavour, $SU(5)_{f}\times SU(5)_{c}$.}. Thus
the magnetic quarks $\varphi$, $\tilde{\varphi}$ decompose into
4 singlets (which we will call $\phi$, $\tilde{\phi}$) plus 2 fundamentals
of $SU(5)_{f}$ (which we call $\rho$, $\tilde{\rho}$), while the
magnetic mesons $\Phi_{ij}$ decompose into 4 fundamentals of $SU(5)_{f}$
($Z$ and $\tilde{Z}$), an adjoint+trace singlet of $SU(5)_{f}$ ($X$),
plus 4 more singlets ($Y$). The charges are as follows:

\begin{center}\begin{tabular}{|c|c|c|c|}
\hline
&
{\small $SU(5)$}&
$SU(2)_{mg}$&
{\small $U(1)_{R}$}\tabularnewline
\hline
\hline
$\Phi_{ij}\equiv\left(\begin{array}{cc}
Y & Z\\
\tilde{Z} & X\end{array}\right)$&
$\left(\begin{array}{cc}
4\times1 & \square\\
\bar{\square} & Adj+1\end{array}\right)$&
$\left(\begin{array}{cc}
1 & 1\\
1 & 1\end{array}\right)$&
2\tabularnewline
\hline
{\small $\varphi\equiv\left(\begin{array}{c}
\phi\\
\rho\end{array}\right)$}&
$\left(\begin{array}{c}
1\\
\bar{\square}\end{array}\right)$&
$\square$&
$1$\tabularnewline
\hline
{\small $\tilde{\varphi}\equiv\left(\begin{array}{c}
\tilde{\phi}\\
\tilde{\rho}\end{array}\right)$}&
$\left(\begin{array}{c}
1\\
\square\end{array}\right)$&
$\bar{\square}$&
$-1$\tabularnewline
\hline
\end{tabular}\par\end{center}

At this point it is worth noting that, thanks to the baryon, the model
has $R$-charges that are not 0 or 2. As discussed in Ref.~\cite{Shih:2007av}
this condition is necessary for Wess-Zumino models
to spontaneously break $R$-symmetry. Therefore, our model allows for spontaneous $R$ symmetry breaking
and we will see in the following that this does indeed happen.

Now let us consider the potential at tree-level. The $F$-term contribution
to the potential at tree-level is
\begin{eqnarray}
V_{F} & = & \sum_{ar}|Y_{rs}\tilde{\phi}_{s}^{a}+Z_{r\hat{i}}\tilde{\rho}_{\hat{i}}^{a}
+2m\varepsilon_{ab}\varepsilon_{rs}\phi_{s}^{b}|^{2}
\\\nonumber
&&\!\!\!\!\!\!+\sum_{a\hat{i}}|\tilde{Z}_{\hat{i}r}\tilde{\phi}_{r}^{a}+X_{\hat{i}\hat{j}}\tilde{\rho}_{\hat{j}}^{a}|^{2}+
\sum_{as}|\phi_{r}^{a}Y_{rs}+\rho_{\hat{i}}^{a}\tilde{Z}_{\hat{i}s}|^{2}
+\sum_{a\hat{j}}|\phi_{r}^{a}Z_{r\hat{j}}+\rho_{\hat{i}}^{a}X_{\hat{i}\hat{j}}|^{2}
\\\nonumber
&&\!\!\!\!\!\!+
 \sum_{rs}|(\phi_{r}.\tilde{\phi_{s}}-\mu^{2}\delta_{rs})|^{2}+\sum_{r\hat{i}}|\phi_{r}.\tilde{\rho}_{\hat{i}}|^{2}
+\sum_{r\hat{i}}|\rho_{\hat{i}}.\tilde{\phi}_{s}|^{2}+\sum_{\hat{i}\hat{j}}|(\rho_{\hat{i}}.\tilde{\rho}_{\hat{j}}-\hat{\mu}^{2}\delta_{\hat{i}\hat{j}})|^{2}
 \end{eqnarray}
where $a,\, b$ are $SU(2)_{mg}$ indices. The flavor indices $r,\,s$ and $\hat{i},\hat{\, j}$  correspond to the $SU(2)_{f}$ and $SU(5)_{f}$, respectively.
It is straightforward to see that the rank condition
works as in ISS; that is the minimum for a given value of $X,Y,Z$
and $\tilde{Z}$ is along $\rho=\tilde{\rho}=0$ and
\begin{eqnarray}
\label{tree1}
\vev{\phi} & = & \frac{\mu^{2}}{\xi}\,\mathbf{I}_{2}\nonumber \\
\vev{\tilde{\phi}} & = & \xi\,\mathbf{I}_{2},\end{eqnarray}
where $\xi$ parameterizes a runaway direction that will eventually
be stabilized by the Coleman-Weinberg contribution to the potential.
This then gives \begin{equation}
Z=\tilde{Z}=0\end{equation}
but the pseudo-Goldstone modes $X=\chi\,\mathbf{I_{5}}$ are undetermined.
(Note that all the $D$-terms are zero along this direction and the
$SU(2)_{mg}$ is Higgsed but $SU(5)_{f}$ is unbroken.) In addition $Y$ becomes
diagonal and real (assuming $m$ is real).
Defining $\vev{Y_{rs}}=\eta\,\mathbf{I}_{2}$, the full potential is
\begin{equation}
\label{potential}
V=2\left|\eta\,\xi+2m\frac{\mu^{2}}{\xi}\right|^{2}+2\left|\eta\frac{\mu^{2}}{\xi}\right|^{2}+5\hat{\mu}^{4}.
\end{equation}
Using $R$ symmetry we can choose $\xi$ to be real\footnote{The phase of $\xi$ corresponds to the $R$-axion which will be dealt with later.}.
Minimizing in $\eta$ we find
\begin{equation}
\label{tree2}
\eta=-2m\left(\frac{\xi^{2}}{\mu^{2}}+\frac{\mu^{2}}{\xi^{2}}\right)^{-1}.
\end{equation}
Substituting $\eta(\xi)$ into Eq. \eqref{potential} we see that $\xi\rightarrow\infty$ is a runaway direction along which \begin{equation}
V(\xi)=8m^{2}\mu^{2}\left(\frac{\xi^{6}}{\mu^{6}}+\frac{\xi^{2}}{\mu^{2}}\right)^{-1}+5\hat{\mu}^{4}.\end{equation}
It is worth emphasizing that even in the limit $\xi\rightarrow\infty$, the scalar potential $V$ is non-zero, so we have a
runaway to \emph{broken} SUSY (a `pseudo-runaway' in the language of \cite{Essig:2007xk}).
Proceeding to one loop, the Coleman-Weinberg contribution to the
potential is therefore expected to lift and stabilize this direction at the same time as lifting the
pseudo-Goldstone modes $\chi$.

Let's see how this works.
Firstly, recall that the Coleman-Weinberg effective potential \cite{Coleman:1973jx} sums up all one-loop quantum
corrections into the following form:

\be
V_{\mathrm{eff}}^{(1)}\!=\!\frac{1}{64\pi^2}\,\mathrm{STr}\,\M^4\log\frac{\M^2}{
\Lambda^2_{UV}}\,
\!\equiv\frac{1}{64\pi^2}\left( \Tr\, m_0^4\log\frac{m_0^2}{\Lambda^2_{UV}}-2\,\Tr\,
  m_{1/2}^4\log\frac{m_{1/2}^2}{\Lambda^2_{UV}} +3\, \Tr\,
m_1^4\log\frac{m_1^2}{\Lambda^2_{UV}}\right)\label{CW}
\ee
where $\Lambda_{UV}$ is the UV cutoff\footnote{As usual we can ``eliminate'' $\Lambda_{UV}$ by trading it for a renormalization scale at which the couplings are defined.},
and the mass matrices are given
by~\cite{Barbieri:1982nz}:

\be
m_0^2=
\begin{pmatrix}
W^{ab}W_{bc}+D^{\alpha a}D^\alpha_{\phantom{\alpha}c}+D^{\alpha
a}_{\phantom{\alpha}c}D^\alpha\quad &
W^{abc}W_b+D^{\alpha a}D^{\alpha c}\\
W_{abc}W^b+D^{\alpha}_{\phantom{\alpha}a}D^{\alpha}_{\phantom{\alpha}c} &
W_{ab}W^{bc}+D^{\alpha}_{\phantom{\alpha}a}D^{\alpha c}+D^{\alpha
c}_{\phantom{\alpha}a}D^\alpha
\end{pmatrix}
\ee

\be
m_{1/2}^2=
\begin{pmatrix}
W^{ab}W_{bc}+2D^{\alpha a}D^\alpha_{\phantom{\alpha}c}\quad
&\quad-\sqrt{2}W^{ab}D^\beta_{\phantom{\beta}b}\\
-\sqrt{2}D^{\alpha b}W_{bc} & 2D^{\alpha c}D^\beta_{\phantom{\beta}c}
\end{pmatrix}
\qquad
m_1^2=D^\alpha_{\phantom{\alpha}a}D^{\beta a}+D^{\alpha
a}D^\beta_{\phantom{\beta}a}.
\ee
As usual, $W_c\equiv \partial W/\partial \Phi^c$ denotes a derivative of the
superpotential with respect to the scalar component of the superfield $\Phi^c$,
and $D^{\alpha}$ are the appropriate $D$-terms, $D^\alpha = g z_a T^{\alpha a}_{\phantom{\alpha}b}z^b$.
Of course, $D$-terms can be switched off by setting the gauge coupling $g=0$, which we will do until further notice.
All the above mass matrices will generally depend on
field expectation values.
The effective potential \mbox{$V_{\rm{eff}}=V_{F}+V^{(1)}_{\rm{eff}}$} is the sum of the F-term (tree-level) and the Coleman-Weinberg contributions.
To find the vacua of the theory we now have to minimize $V_{\rm{eff}}$. The true,
stable vacuum will be the global minimum, with other
minima being only meta-stable.

It is interesting to note that as the 1-loop corrections are of a supertrace
form, they vanish around supersymmetric vacua. If the runaway was to a
supersymmetric vacuum at infinity, the Coleman-Weinberg corrections wouldn't lift it. In
our case, we have a runaway to a non-supersymmetric vacuum at infinity, so it
is reasonable to expect that these loop corrections will modify the asymptotic
behaviour.

Now let's see how the classical runaway direction is lifted by quantum
effects. We parameterize the pseudo-Goldstone and runaway field vacuum expectation values by
\bea
\label{phivevs}
\vev{\tilde\phi}\!\!\!&=&\!\!\!\xi\,\mathbf{I}_{2}\quad\quad\quad\quad\,\,\vev{\phi}=\kappa\,\mathbf{I}_{2}\\
\label{yvevs}
\vev{Y}\!\!\!&=&\!\!\!\eta\,\mathbf{I}_{2}\quad\quad\quad\quad\vev{X}=\chi\,\mathbf{I}_{5}.
\eea
These are the most general vevs consistent with the tree level minimization.
It can be checked that at one loop order all other field vevs are zero in the lowest perturbative vacuum. By computing the masses of all fluctuations about this valley we can go about
constructing the one-loop effective potential from eq. \eqref{CW}.
We have done this numerically using {\em{Mathematica}} as well as {\em{Vscape}}, a program specifically
written to explore the properties of metastable vacua~\cite{van den Broek:2007kj}.

Table \ref{tabvev} shows the result of minimizing the vevs in the one-loop effective potential for some sample values of the parameters. As expected, the
vevs in Eqs. \eqref{phivevs}, \eqref{yvevs} are seen to approximately, i.e. up to small Coleman-Weinberg corrections, satisfy the analytic
tree-level relations \eqref{tree1}, \eqref{tree2}
\be
\kappa=\frac{\mu^2}{\xi}\quad\quad\quad\eta=-2m\left(\frac{\xi^{2}}{\mu^{2}}+\frac{\mu^{2}}{\xi^{2}}\right)^{-1}.
\ee
We have checked that this is indeed the case for a wide range of input parameters.
Hence, in what follows, we impose the condition above and only consider the two independent vevs $\xi$ and $\chi$.

\begin{figure}[t]
\begin{center}
\begin{picture}(300,200)
\Text(-25,170)[l]{\scalebox{1.3}[1.3]{$V/\hat{\mu}^{4}$}}
\Text(265,-8)[l]{\scalebox{1.3}[1.3]{$\xi/\hat{\mu}$}}
\includegraphics*[bbllx=91,bblly=3,bburx=493,bbury=251,width=.65\textwidth]{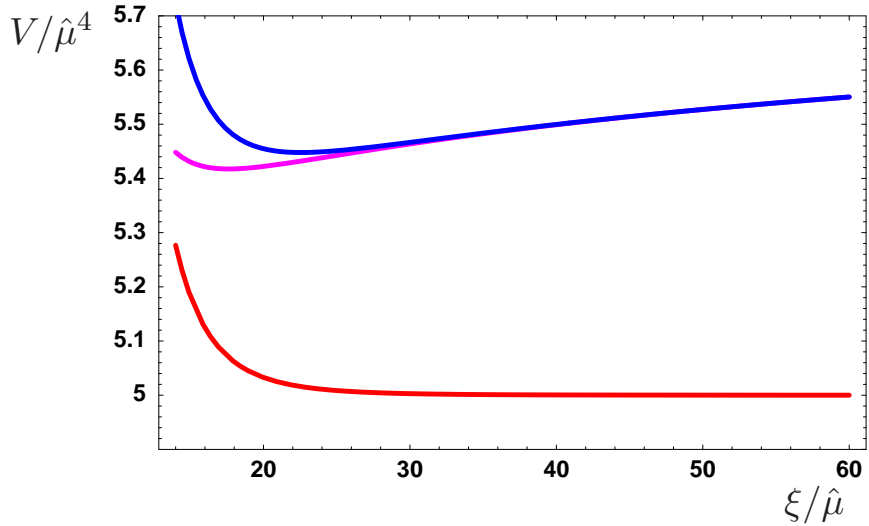}
\end{picture}
\end{center}
\caption{\small This plot demonstrates the stabilization in the $\xi$ direction. The red curve is the tree-level runaway potential.
The purple is the Coleman-Weinberg contribution (we have added a constant shift of 5 to it). The blue line
depicts the full stabilized potential. (We use $\mu=4\hat{\mu}$, $m=2\hat{\mu}$.)}\label{xiplot}
\end{figure}

\begin{figure}[t]
\begin{center}
\begin{picture}(300,300)
\Text(-10,153)[l]{\scalebox{1.3}[1.3]{$\chi/\hat{\mu}$}}
\Text(160,0)[l]{\scalebox{1.3}[1.3]{$\xi/\hat{\mu}$}}
\includegraphics*[bbllx=91,bblly=3,bburx=430,bbury=341,width=.65\textwidth]{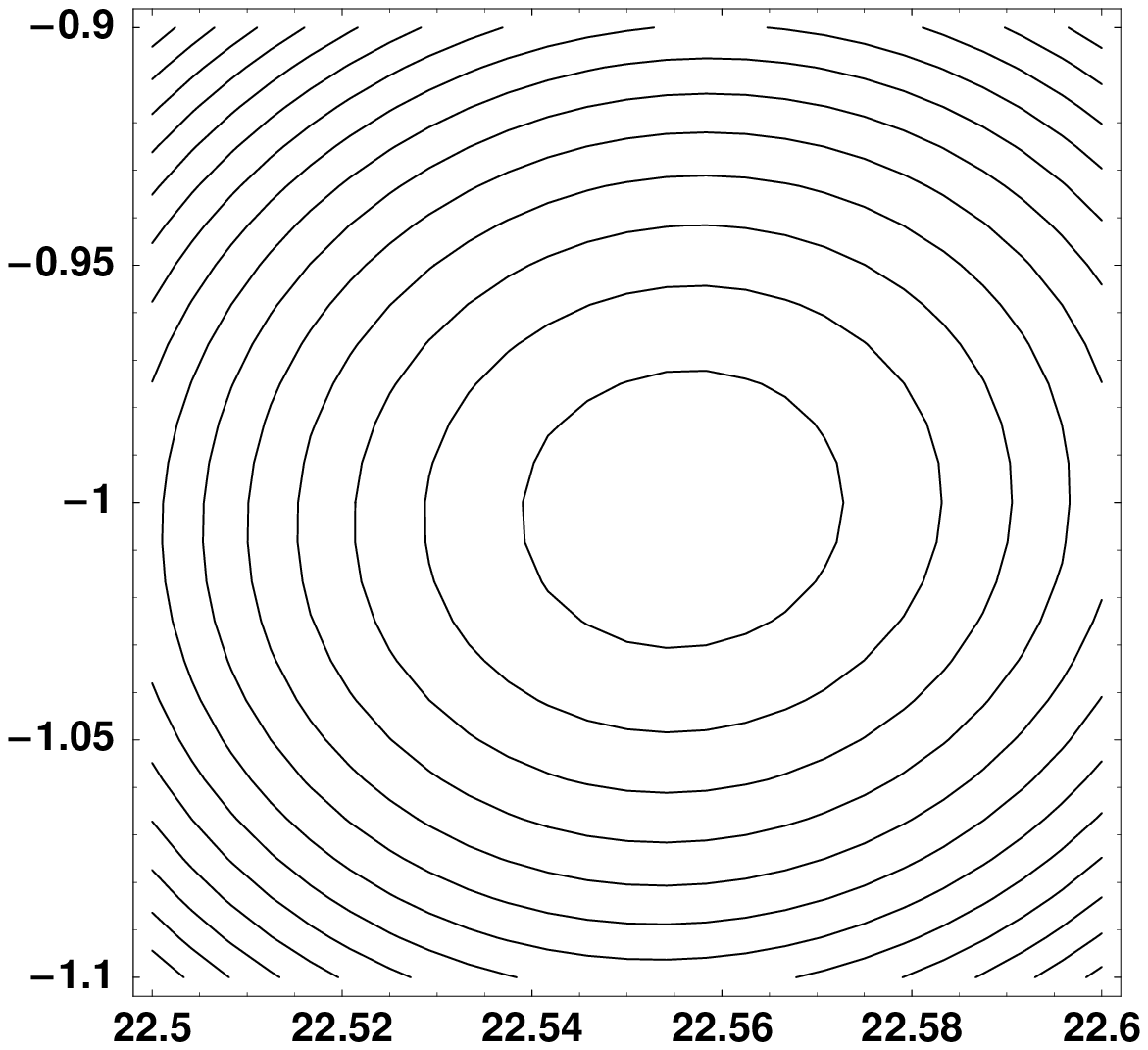}
\end{picture}
\end{center}
\caption{\small This contour plot of the effective potential $V_{\rm{eff}}$ shows that the pseudo-modulus $\chi$ is also stabilized
at a non-vanishing vev. (We use $\mu=4\hat{\mu}$, $m=2\hat{\mu}$.)}\label{contourplot}
\end{figure}

A plot of the potential in the $\xi$ direction, Fig. \ref{xiplot}, shows the Coleman-Weinberg terms do indeed stabilize
the $\xi \rightarrow \infty$ runaway at finite, non-zero values of the fields. A contour plot in the $\xi-\chi$ plane, Fig. \ref{contourplot},
reveals that the pseudomodulus $\chi$ is also stabilized at a non-zero value $\mathcal{O}(\hat{\mu})$.

\begin{table}[t]
\begin{center}
\begin{tabular}{|c|c|c|c|c|}
\hline
Model &  $\xi/\hat{\mu}$ & $\kappa/\hat{\mu}$ & $\eta/\hat{\mu}$& $\chi/\hat{\mu}$\tabularnewline
\hline
\hline
Vscape Unconstrained & 22.55451 & 0.709338  & $-0.125660$ &
$-1.00041$ \tabularnewline
Vscape Constrained & 22.55581 & $0.709352^\dag$ & $-0.125671^\dag$ & $-1.00132$ \tabularnewline
Mathematica & 22.5559 & $0.70935^\dag$ & $-0.12567^\dag$ & $-1.0014$ \tabularnewline
Gauged $SU(2)_{mg}$, $g=0.4$ & 22.4385 & $0.71306^\dag$ & $-0.12699^\dag$ & $-1.0115$ \tabularnewline
\hline \hline
\end{tabular}\par
\end{center}
\caption{Stabilized vevs for different models: $\mu=4\hat{\mu}$, $m=2\hat{\mu}$. The values$^\dag$ are obtained from the tree-level
constraints Eqs. \eqref{tree1}, \eqref{tree2}.}\label{tabvev}
\end{table}

Thus, for a natural choice of parameters, all the vevs $\xi$, $\kappa$, $\eta$ and $\chi$ obtain stable, finite $\mathcal{O}(\hat{\mu})$ values.
Notice that $\Phi$, $\varphi$ and $\tilde\varphi$ all carry $R$-charge, so the $R$-symmetry of the model is spontaneously broken in this minimum.

Until now we have neglected the $D$-terms from $SU(2)_{mg}$ but, as we can see from Tab. \ref{tabvev}, including them does not significantly alter the vev-structure
of the vacuum.

What about the stability of this vacuum? When the gauge fields are turned on, this model has non-zero Witten index, so the global minimum will be supersymmetric.
As in the ISS model, this minimum is induced by the non-perturbative contribution to the superpotential,
\begin{equation}
W_{np}=2\Lambda^{3}\left[\det\left(\frac{\Phi}{\Lambda}\right)\right]^{\frac{1}{2}}.
\end{equation}
Adapting the supersymmetric vacuum solution from the ISS model to our case with $\mu>\hat{\mu}$ we find,
\begin{equation}
\varphi=0,\quad\tilde{\varphi}=0,\quad\eta=\hat{\mu}^{2}\mu^{-\frac{6}{5}}\Lambda^{\frac{1}{5}}\quad \chi=\mu^{\frac{4}{5}}\Lambda^{\frac{1}{5}}.
\end{equation}
Note that the supersymmetric minimum lies at $\varphi=\tilde{\varphi}=0$ and is completely
unaffected by the baryon
deformation.
\medskip

So far we have established that supersymmetry is broken dynamically
and $R$-symmetry spontaneously in the metastable vacuum of the ISS
sector. We now need to transmit both these effects to the Standard
Model. The most concise way to do it is to gauge the $SU(5)_{f}$ and
to identify it with the parent gauge group the Standard Model. Since
both supersymmetry and $R$-symmetry are broken\footnote{In contrast to the ISS model
which only has small anomalous R-symmetry breaking our model
has in addition a rather large spontaneous R-symmetry breaking by the vacuum expectation
value $\langle \chi\rangle$.}, gauginos do
acquire a mass.

{
Gaugino masses are generated at one loop order (cf. Fig. \ref{gauginofig}). The fields propagating in the loop are
fermion and scalar components of the direct mediation `messengers' $\rho$, $\tilde{\rho}$ and $Z$, $\tilde{Z}$.
For fermion components,
\begin{equation}
\psi =  (\rho_{ia}\,,Z_{ir})_{ferm} , \quad
\tilde{\psi}  =  (\tilde{\rho}_{ia}\,,\tilde{Z}_{ir})_{ferm},
\end{equation}
the mass matrix is given by
\begin{equation}
m_{f}=\mathbf{I}_{5}\otimes\mathbf{I}_{2}\otimes\left(\begin{array}{cc}
\chi & \xi\\
\frac{\mu^{2}}{\xi} & 0\end{array}\right).
\label{mferms}
\end{equation}
We assemble the relevant scalars into
\begin{equation}
(\rho_{ia},Z_{ir},\tilde{\rho}_{ia}^{*},\tilde{Z}_{ir}^{*})_{sc},
\end{equation}
and for the corresponding scalar mass-squared matrix we have
\begin{equation} \label{m2sc}
m_{sc}^{2}=\mathbf{I}_{5}\otimes\mathbf{I}_{2}\otimes\left(\begin{array}{cccc}
|\xi|^{2}+|\chi|^{2} & \chi\frac{|\mu|²}{\xi^{*}} & -\hat{\mu}^{2} & \eta{}^{*}\frac{|\mu|²}{\xi^{*}}\\
\chi\frac{|\mu|²}{\xi} & \frac{|\mu|^{4}}{|\xi|^{2}} & (\xi\eta)^{*}+2m\frac{|{\mu|}^{2}}{\xi} & 0\\
-\hat{\mu}^{2} & \xi\eta+2m\frac{|{\mu|}^{2}}{\xi^{*}} & \frac{|\mu|^{4}}{|\xi|^{2}}+|\chi|^{2} & (\chi{\xi)}^{*}\\
\eta\frac{|\mu|²}{\xi} & 0 & \chi\xi & |\xi|^{2}\end{array}\right).
\end{equation}

Gaugino masses arise from the one-loop diagram in Fig. \ref{gauginofig}. Due to the non-diagonal form of the
matrices \eqref{mferms}, \eqref{m2sc},
we find it easiest to evaluate the appropriate expressions numerically.
The scale $\hat\mu$ is the SUSY-breaking scale. We will keep it fixed,
and measure all other dimensionful parameters in units of $\hat\mu$.
Then for fixed $\hat\mu=1$ there are only two independent input parameters, $\mu$ and $m$,
while the VEVs $\xi$, $\kappa$, $\eta$ and $\chi$ are generated from minimizing the effective potential,
as above. For the purposes of this paper
we will focus on generating the largest possible values for
gaugino masses (in units of $\hat\mu$).\footnote{In Ref.~\cite{SCJV} we will explore the parameter space
of the model in more detail.}
We find that this occurs when $\mu \simeq \hat\mu.$
For example, for $\mu = 1.1\, \hat\mu$ and $m= 0.3 \,\hat\mu$ we have
\begin{equation}
m_{\lambda_{A}}  \simeq  \frac{g_{A}^{2}}{16\pi^{2}} \, 0.0089 \, \hat\mu,
\label{mlnum}
\end{equation}
where $A=1,2,3$ labels the three gauge groups of the Standard Model.
Requiring that all the gaugino masses are
\begin{equation}
m_{\lambda_{A}}\sim (0.1-1)\,\rm{TeV},
\end{equation}
we conclude that
\begin{equation}
\hat{\mu}\sim (10^4-10^5)\,{\rm{TeV}}
\end{equation}
in this point in the parameter space of our model.

We would like to compare this numerical evaluation of the gaugino mass with
the simple analytical expression one might have anticipated.
Assuming that
the dominant effect comes from magnetic quarks,
$\rho$ and $\tilde{\rho}$, propagating in the loop, as shown in Fig.
\ref{gauginofig}, and working to the leading order in susy breaking, i.e. to order $F_{\chi}$,
gaugino mass goes as
\begin{eqnarray}
\label{gauginowrong}
m^{\rm{naive\,\,estimate}}_{\lambda_{A}}  \sim  \frac{g_{A}^{2}}{16\pi^{2}}\frac{\langle F_{\chi}\rangle}{\langle\chi\rangle}
\sim  \frac{g_{A}^{2}}{16\pi^{2}}\frac{\hat{\mu}^2}{\langle\chi\rangle} \sim  \frac{g_{A}^{2}}{16\pi^{2}} \hat\mu.
 \end{eqnarray}
For the last part of \eqref{gauginowrong}
we have assumed that all VEVs and mass parameters are of the same order ${\cal O}  (\hat\mu).$
This expression is two orders of magnitude greater than our
numerical result and therefore is too simplistic to give a correct estimate.
In fact, the correct ${\cal O}(F_{\chi})$
contribution to $m^{(1)}_{\lambda_{A}} \sim \frac{g_{A}^{2}}{16\pi^{2}}\, F_{\chi}\,
(m_{f})^{-1}_{11}$ vanishes in our model, and one needs to go to order $F_{\chi}^3$ to find
a non-vanishing contribution. This effect was first pointed out in Ref.~\cite{Izawa:1997gs}:
the zero
element in the lower right corner of the fermion mass matrix
\eqref{mferms} implies that $(m_{f})^{-1}_{11} =0$ and hence $m^{(1)}_{\lambda_{A}}=0$.

The vanishing of the leading order contribution
$m^{(1)}_{\lambda_{A}}$ to the gaugino mass contrasts with the usual
gauge mediation argument that the scalar masses should be roughly
similar to the gaugino masses $m_{\lambda_{A}}\sim M_{sc}$. Clearly,
the $R$-symmetry breaking (together with the structure of the
messenger mass matrices) plays a crucial role in suppressing the
gaugino masses.
Since scalars are not protected by R-symmetry the generation of their masses is less constrained\footnote{For example, as long as supersymmetry is broken, we can have
scalar masses even when R-symmetry is unbroken.}.
Hence, we
expect the appropriate two-loop diagrams
(cf., e.g., \cite{Martin:1996zb,Giudice:1998bp}) to
give something closely approximating the naive estimate for the scalar
masses,
\begin{equation}
M_{sc}^2 \sim (\frac{g_{A}^{2}}{16\pi^{2}})^2 \, \hat\mu^2 \, .
\end{equation}

In our model therefore the scalars are
always heavier than the gauginos.
The phenomenology for this particular type of model is
expected to be of the "heavy-scalar" type as reviewed in
Ref.~\cite{heavyscalar}. In the region $\hat{\mu}\simeq \mu
\sim m$ their masses are only about two orders of magnitude larger
than the gaugino masses, and a focus-point type of phenomenology
\cite{Feng:1999zg} may be possible. (Note that by choosing $\mu$ to
be even closer to $\hat{\mu}$ one may get gaugino masses closer to
those of the scalars.) Increasing $\mu$ and decreasing $m$ takes us
continuously to the split SUSY scenario
\cite{ArkaniHamed:2004fb,Giudice:2004tc}. A fuller investigation
will be carried out in Ref.~\cite{SCJV}.

Non-perturbative effects due to $W_{\rm{np}}$ are suppressed by the scale $\Lambda$ of the Landau pole of the ISS sector, which we have not yet constrained. Choosing
$\Lambda\gg\hat{\mu}$ (so that the magnetic theory is weakly coupled and the metastable vacuum is long lived) the non-perturbative corrections to our discussion
are small.

\begin{figure}[t]
\begin{center}
\begin{picture}(200,120)
\includegraphics*[bbllx=143,bblly=549,bburx=445,bbury=773,width=.35\textwidth]{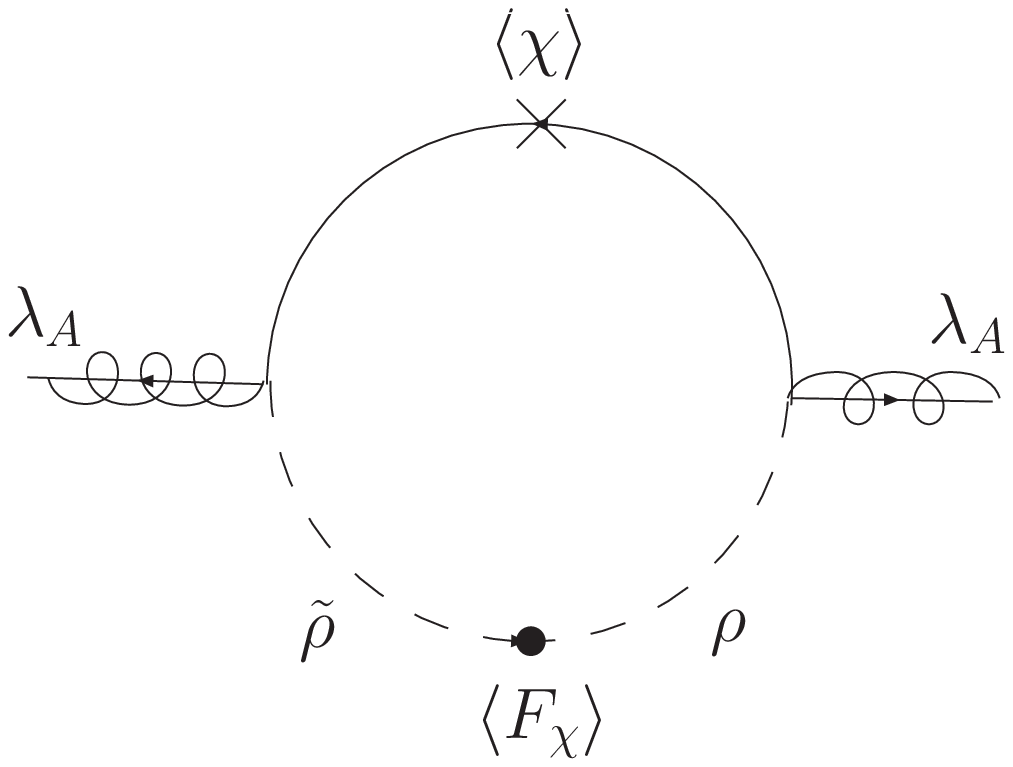}
\end{picture}
\end{center}
\vspace*{-0.5cm}
\caption{\small One-loop contribution to the gaugino masses. The blob on the scalar line indicates
an appropriate number of insertions of $\langle F_{\chi}\rangle$ to make the diagram non-vanishing.}\label{gauginofig}
\end{figure}
\medskip

Our model has a spontaneously broken R-symmetry that is explicitly broken
only by the non-perturbative contribution $W_{np}$ to the superpotential.
In such a situation we generally expect a pseudo-Goldstone boson --
the R-axion $a_{R}$ (cf, e.g.,  \cite{Dine:1993yw,Bagger:1994hh,Nelson:1993nf}). If such a particle is light it can have dangerous
phenomenological consequences \cite{Kim:1986ax,Raffelt:2006cw,Raffelt:1996}. Since the R-symmetry is an axial symmetry
triangle diagrams typically couple the R-axion to gauge fields via
a term (see, e.g., \cite{Kim:1986ax})
\begin{equation}
\sim\frac{\alpha}{2\pi f_{R}}a_{R}\,F^{\mu\nu}\tilde{F}_{\mu\nu}
\end{equation}
 where $F^{\mu\nu}$ is a gauge field and $f_{R}$ is the scale of spontaneous
R-symmetry breaking. Particularly dangerous are the couplings of this
type to gluons and photons. Moreover, there can exist couplings of the $R$-axion to matter fields.
For small masses \mbox{$m_{a_{R}}\lesssim100\,{\rm {MeV}}$}
astrophysical considerations \cite{Raffelt:2006cw,Raffelt:1996} constrain the scale of spontaneous
R-symmetry breaking to be
\begin{equation}
f_{R}\gtrsim{\rm{few}}\times 10^{7}\,{\rm {GeV}}\quad{\rm {for}}\quad m_{a_{R}}\lesssim100\,{\rm {MeV}}.
\end{equation}

Let us now estimate the mass of the R-axion in our model to check
whether it is harmless. The R-axion is the phase of the fields that
spontaneously break the R-symmetry,
\begin{equation}
\eta=|\eta|\exp\left(2{\rm {i}}\frac{a_{R}}{f_{R}}\right),\quad\chi=|\chi|\exp\left(2{\rm {i}}\frac{a_{R}}{f_{R}}\right),\label{fields}
\end{equation}
 where the $2$ arises from the R-charge $2$ of the $\Phi$-field.
The dominant contribution to spontaneous R-symmetry breaking comes
from $\langle\eta\rangle$. This sets the scale \begin{equation}
f_{R}\sim\langle\eta\rangle.\end{equation}
 The R-axion mass arises from the explicit breaking due to%
\footnote{Another contribution to the R-axion mass may arise from supergravity.
A constant term in the superpotential that cancels the cosmological
constant also breaks R-symmetry explicitly \cite{Bagger:1994hh}.%
} $W_{np}$.
More precisely, taking into account $W_{np}$ contribution to the $F_{X}$-terms,
\begin{eqnarray}
V_{F}\ni|F_{X}|^{2}\!\! & \sim & \!\!\left|\langle\eta\rangle
\langle\chi\rangle^{\frac{3}{2}}\exp\left(5{\rm {i}}\frac{a_{R}}{\langle\eta\rangle}\right)\Lambda^{-\frac{1}{2}}-\hat{\mu}^{2}\right|^{2}\label{potential2}\\
\!\! & = &
\!\!\left[\langle\eta\rangle^{2}\langle\chi\rangle^{3}\Lambda^{-1}+\hat{\mu}^{4}
-2\hat{\mu}^{2}\langle\eta\rangle\langle\chi\rangle^{\frac{3}{2}}
\Lambda^{-\frac{1}{2}}\cos\left(5\frac{a_{R}}{\langle\eta\rangle}\right)\right],\nonumber
\end{eqnarray}
the $R$-axion mass arises from the last term on the right hand side.
(For simplicity, we have chosen $\hat{\mu}$ and all the vevs
to be real.)
Expanding to second order in $a_{R}$ we find the R-axion
mass to be, \begin{equation}
m_{a_{R}}^{2}\sim \hat{\mu}^{2}\langle\eta\rangle^{-1}\langle\chi\rangle^{\frac{3}{2}}\Lambda^{-\frac{1}{2}}.\end{equation}
 For our values this turns out to be sufficiently heavy to easily avoid the astrophysical constraints for any $\Lambda<M_{P}$.
\medskip

We now want to comment on a particular feature of our model, and indeed
all direct mediation models based on embedding the Standard Model gauge group
into the flavor group of the ISS sector.
As already mentioned in refs.\cite{ISS,Csaki:2006wi} this embedding adds a significant number of
matter multiplets charged under the SM gauge groups. Above the mass thresholds of these fields this leads to
all Standard Model gauge groups being not asymptotically free and therefore to Landau poles in the SM sector.
Since the additional fields are in $SU(5)$ multiplets, the beta functions
of the SM gauge couplings are modified universally. For example, in our model as
\begin{equation}
b_{A}=b_{A}^{(MSSM)}-9
\end{equation}
where the additional contributions are $2$ from $\varphi$ and $\tilde{\varphi}$,
and 7 from $\Phi$. The SM gauge couplings at a scale $Q>\mu$ in
our model are therefore related to the traditional MSSM ones as
\begin{equation}
\alpha_{A}^{-1}=(\alpha_{A}^{-1})^{(MSSM)}-\frac{9}{2\pi}\log(Q/\hat{\mu}),
\end{equation}
where the fields $\varphi,$ $\tilde{\varphi}$ and $\Phi$ contribute
to the running above the scale $\mu$. The Landau pole $Q\equiv\Lambda^{(MSSM)}$
we will take to be situated where $g_{A}\sim4\pi$ which corresponds
roughly to
\begin{equation}
\frac{\Lambda^{(MSSM)}}{\hat{\mu}}\sim 10^{5}.
\end{equation}
Values of $\hat{\mu}\gtrsim 10^{8}TeV$ would be required in order to reach
the conventional GUT scale in the MSSM sector before the Landau pole.

We would now like to suggest that the change of sign in the slopes of the Standard Model gauge couplings
and the very existence of Landau poles is an interesting feature rather than an insurmountable problem.
Presence of Landau poles in all sectors of theory indicates that we should interpret not only the ISS sector as a magnetic dual of an asymptotically free theory, but also
apply the same reasoning to the Standard Model itself.
In other words, at energy scales above $\hat{\mu}$ the Standard Model sector and the ISS sector are not decoupled from each
other and, in general, should be treated as part of the same theory.
We already know that the UV completion of the ISS sector is its electric Seiberg dual and we propose the whole theory has such a UV completion.
This seems to be a rather symmetric construction.
One consequence of this interpretation is, of course,
that gauge unification is lost, or at least buried in the unknown details of the dual theory.

{\em{In summary}} we find that direct mediation (i.e. mediation
in which there is no separate messenger sector) is relatively simple
to implement in ISS-like models by inducing spontaneous breakdown
of the approximate $R$-symmetry associated with the metastable minimum which in turn allows us to generate gaugino masses alongside other soft SUSY-breaking terms.
We presented a baryon-deformed ISS model in which this occurs automatically
due to the Coleman-Weinberg potential. Once the $R$-symmetry is broken,
the magnetic quarks of the ISS sector are able to play the role of
messengers, if one identifies an $SU(5)_{f}$ subset of the flavour
symmetry with the SM gauge groups. We have speculated that the entire theory is a magnetic dual of an
(unknown) asymptotically free theory.

\end{document}